\documentclass[12pt]{article}

\def\be{\begin{equation}}
\def\ee{\end{equation}}
\def\ba{\begin{array}{c}}
\def\baa{\begin{array}{ll}}
\def\ea{\end{array}}

\def\ben{$$}
\def\een{$$}
 
\begin{document}

\titlepage

  \begin{center}{\Large \bf

Discrete ${\cal PT}-$symmetric models of scattering

 }\end{center}

\vspace{5mm}

  \begin{center}

{\bf Miloslav Znojil}

 \vspace{3mm}

Nuclear Physics Institute ASCR, 250 68 \v{R}e\v{z}, Czech
Republic\footnote{ e-mail: znojil@ujf.cas.cz}\\

 \vspace{3mm}

\end{center}

\vspace{5mm}

\section*{Abstract}

One-dimensional scattering mediated by non-Hermitian Hamiltonians
is studied. A schematic set of models is used which simulate two
point interactions at a variable strength and distance. The
feasibility of the exact construction of the amplitudes is
achieved via the discretization of the coordinate. By direct
construction it is shown that in all our models the probability is
conserved. This feature is tentatively attributed to the space-
and time-reflection symmetry (a.k.a. ${\cal PT}-$symmetry) of our
specific Hamiltonians.

 \vspace{9mm}

\noindent PACS  03.65.Nk, 03.80.+r, 11.55.Ds, 03.65.Ge,

\noindent MSC 2000: 81U15, 81Q05, 81Q10, 46C20, 47B36, 39A70




\section{Introduction}

In the absence of an external potential, the motion of a quantum
particle is described by the kinetic-energy Hamiltonian $H_0 =-
d^2/dx^2$ in one dimension ($\hbar=2m=1$). This operator is
Hermitian and, incidentally, symmetric with respect to the space
and time reflection (i.e., ${\cal PT}-$symmetric, $H_0{\cal
PT}={\cal PT}H_0$, cf. many relevant comments on such a type of
symmetry in \cite{Carl}).

In an approximation where the real line is replaced by the mere
discrete lattice of coordinates with some sufficiently small
stepsize $h>0$,
 \ben
  x_k=k\,h\,,\ \ \ \ \
 \ \ \ \ \ k = 0, \pm 1, \ldots\,
 \een
the role of the kinetic energy is often being played by the doubly
infinite tridiagonal matrices
 \ben
 H_0'=
 \left [\begin {array}{rrrrr}
  \ddots&\ddots&&&\\{}
  \ddots&2&-1&&
 \\{}
 &-1&2&-1&\\{}&&-1&2&\ddots
 \\{}
 &&&\ddots&\ddots\end {array}\right ]\,
 \ \ \ \ \ {\rm or}\ \ \ \ \
 H_0=
 \left [\begin {array}{rrrrr}
  \ddots&\ddots&&&\\{}
  \ddots&0&-1&&
 \\{}
 &-1&0&-1&\\{}&&-1&0&\ddots
 \\{}
 &&&\ddots&\ddots\end {array}\right ]\,
 \een
which differ just by a trivial shift of the energy scale. Whenever
we treat ${\cal P}$ as the parity (${\cal P}x_k=x_{-k}$) and the
antilinear  operator ${\cal T}$ as the time reversal (i.e., in our
present setting, transposition plus complex conjugation), we may
represent the product-operator symmetry of our real matrices $H_0$
by the antidiagonal unit matrix
 \be
 {\cal PT}=
 \left [\begin {array}{rlcrr}
  &&&&\ \ \dot{ \dot{\dot{}\   }\  }\\{}
  &&&\  1 &
 \\{}
 && 1 &&\\{}&1 \ &&&
 \\{}\ \ \ \
\dot{ \dot{\dot{}\   } \ }
 &&&&\end {array}\right ]\,.
 \label{PTdef}
 \ee
Using this definition we shall demand that also all the
nontrivial, doubly infinite discrete  Hamiltonians $H=H_0+W$
possessing a nonvanishing interaction term $W$ will be required
real and ${\cal PT}-$symmetric.

The matrix dimension of the interaction matrix $W$ (i.e., the
``range" of the interaction) will be assumed finite. One expects
that then the scattered states could stay asymptotically
undistorted. In the mathematical terminology such an expectation
means that we feel allowed to search for the solutions of the
discrete and ${\cal PT}-$symmetric Schr\"{o}dinger equations
 \be
 (H_0+W)\,\psi = E\,\psi
 \label{SE}
 \ee
complemented by the {\em standard, undistorted} boundary
conditions
 \be
 \psi_m =
 \left \{
 \begin{array}{ll}
 e^{i\,m\,\varphi}+R\,e^{-i\,m\,\varphi}\,,\ \ \ \ \
 \ \ \ \ \ \
 & m \leq -M\,,\\
 T\,e^{i\,m\,\varphi}\,,\ \ \ \ \ &
 m \geq M\,.
 \ea
 \right .
 \label{discatbc}
 \ee
We should remind the readers that the standard re-parametrization
of the energy $E= (2 -2 \cos \varphi )/h^2$ in terms of the real
angle $\varphi\in (0,\pi)$ should be used \cite{jadis}.

Our study has been inspired by a few papers on the scattering in
non-Hermitian scenario \cite{Ahmed,tobo,Cannata} and, in
particular, by the Jones' paper \cite{Jones}. Unfortunately, its
author worked in the differential-equation limit $h \to 0$ which
made the detailed analysis perceivably hindered by the
non-Hermiticity of the equations. In effect, the feasibility
requirements (cf. \cite{delta}) restricted his attention to the
mere ${\cal PT}-$asymmetric delta-function interactions,
therefore.

In our subsequent comment \cite{prd} we facilitated the
technicalities by the transition to the discretized
eq.~(\ref{SE}). Having preserved the Jones' philosophy we choose
just the ${\cal PT}-$asymmetric models exemplified by the
``ultralocal", two-by-two matrix example
 \ben
 W^{(UL)}=
 \left (
 \begin{array}{cc}
 0&-a\\a&0
 \ea
 \right )\,
 \een
such that $W^{(UL)}\,{\cal PT} \neq {\cal PT}\,W^{(UL)}$. Due to
the discretization approximation $h>0$ we were able to construct
the explicit formulae for the reflection and transmission
coefficients $R$ and $T$, respectively,
 \ben
 R^{(UL)}=-\frac{a^2}{\triangle}\,,
 \ \ \ \ \
 T^{(UL)}=\frac{(1-a)(1- e^{2{\rm i}\varphi})}{\triangle}\,,
 \ \ \ \
 \triangle = 1-(1-a^2)\, e^{2{\rm i}\varphi}\,.
 \een
We were also able to mimic the key features of the Jones'
first-order perturbation results by another entirely exact and
compact formula
 \ben
 \left |R^{(UL)}\right |^2+
 \left |T^{(UL)}\right |^2=\frac{1-{a\,}{[1+
 U(a,\varphi)]^{-1}}}{1+{a\,}{[1+U(a,\varphi)]^{-1}}}
 \,,\ \ \ \ \ \
  U(a,\varphi)=\frac{a^4}{2\,(1-a)\,(1-\cos 2  \varphi)}\,.
  \label{uveled}
 \een
This formula closely resembled the very similar Jones'
perturbation results \cite{Jones}. Hence, we could also parallel
his conclusion that since the predicted sum appears greater than 1
or less than 1 (depending on the sign of the coupling $a$) it
cannot be given the usual probabilistic interpretation. One must
rather assume the presence of some respective ``unknown source" or
``unknown absorber" near the origin. Thus, in the effective-theory
manner, the mathematical non-Hermiticity of the interaction terms
$W$ {\em precisely reflects} the presence of certain hidden
physical mechanisms which violate the conservation of the number
of particles.

In the context of the internal physical consistency of many
non-Hermitian bound-state models \cite{Carl} such an
effective-theory physical interpretation of the scattering looks
rather unsatisfactory. In what follows, for this reason, we shall
try to re-install the ${\cal PT}-$symmetry in our matrix model(s)
and study the consequences. For this purpose we shall make use of
the enhancement of the feasibility of the calculations at a finite
$h>0$. This will make us  able to show that the return to the
simplest ${\cal PT}-$symmetric discrete models finds its
unexpected reward in a {\em complete} suppression and elimination
of the ``unknown" annihilation and creation processes. In the
other words we shall reinstall a firmer parallel between a
simplifying role of ${\cal PT}-$symmetry in {\em both} the
bound-state and scattering-state hypothetical experimental
arrangements.

\section{Solvable discrete models of scattering}

Let us consider the Hamiltonian $ H=H^{(M)}(g)=H_0+W(g)$ of the
doubly infinite matrix form where the non-vanishing part of the
matrix $W(g)=g\,V^{(M)}$ will be linear in the real coupling $g$
and where the matrix $V^{(M)}$ itself will be tridiagonal and
formed just by the four off-diagonal nonvanishing matrix elements.
These elements will be arranged in such a way that using the
definition (\ref{PTdef}), the ${\cal PT}-$symmetry of the complete
Hamiltonian will be guaranteed,
 \be
 V_{1-M,-M}^{(M)}=V_{M-1,M}^{(M)}=1\,,
 \ \ \ \ \ \ V_{-M,1-M}^{(M)}=V_{M,M-1}^{(M)}=-1\,.
 \label{model}
 \ee
The resulting Hamiltonian $H$ can be interpreted as a discrete
kinetic-energy operator complemented by an interaction mimicking
the ${\cal PT}-$symmetrized pair of delta functions \cite{Jakub}.
At the smallest ``distances" $M=1, 2, \ldots$ our model
(\ref{model}) may also resemble certain solvable short-range
square-well differential-operator Hamiltonians \cite{Quesne}. In
the free-motion case the above-mentioned connection between our
$H(0)=H_0$ and the Runge-Kutta Laplacean may be recalled to
explain the origin of the constraint $E \in (0, 4/h^2)$. This is a
peculiarity which is well known in the bound-state context
\cite{jadis}. Here this restriction proves equally important for
the physical consistency of the scattering boundary conditions
(\ref{discatbc}).

In what follows, we intend to search for the solutions of
Schr\"{o}dinger eq.~(\ref{SE}) + (\ref{discatbc}) using the
standard matching method. We should emphasize that in the
scattering scenario the key specific feature of wave functions is
that they are constructed at any energy (from the allowed interval
with, say, $\varphi\in (0,\pi)$) and that they are {\em not}
${\cal PT}-$symmetric themselves (this symmetry is broken by the
boundary conditions). At the same time, due to the compact nature
of the range of our interactions $W$, the non-compact character of
the wave functions is fully characterized by eq.~
(\ref{discatbc}). Thus, in place of the doubly infinite matrix
$H(x)$ with the structure
 \ben
 \left [\begin {array}{cc|cc|cc|cc|cc}
 \noalign{\medskip}{}&\ddots&{}&{}&{}&{}&{}&{}&{}&{}{}\\
 \noalign{\medskip}{}\ddots&{}& -1 &{}&{}&{}&{}&{}&{}&{}{}\\
 \hline
 \noalign{\medskip}{}{}&-1&{}&-1-x&{}&{}&{}&{}&{}&{} \\
 \noalign{\medskip}{}{}&{}&-1+x&{}&-1&{}&{}&{}&{}&{}{}\\
 \hline
 \noalign{\medskip}{}{}&{}&{} &-1&{}&\ddots&{}&{}&{}&{}{}\\
 \noalign{\medskip}{}{}&{}&{} &&{\ddots}&&{-1}&{}&{}&{}{}\\
 \hline
 \noalign{\medskip}{}&{}&{}&{}&{}&-1&{}&-1+x&{}&{}{} \\
 \noalign{\medskip}{}&{}&{}&{}&{}&{}&-1-x&{}&-1&{}{}{}\\
 \hline
 \noalign{\medskip}{}&{}&{}  &{}&{}&{}&{}&-1&{}&\ddots{}{}\\
 \noalign{\medskip}{}&{}&{}&{}&{}&{}&{}&{}&\ddots&{}\end {array}\right ]
 \een
we only have to study the ``central" submatrices of $H$ in which
$W \neq 0$.

In principle, we could consider {\em both} the even- and
odd-dimensional $W$s. Nevertheless, in the context of bound states
we already saw that the difference between the $2M-$ and
$2M+1-$dimensional cases is purely formal \cite{chain}. For this
reason we shall work just with odd dimensions here. This choice
has the two marginal formal merits in containing the ``first
nontrivial" three-dimensional model at $M=1$,
 \ben
 V^{(1)}=\left (
 \begin{array}{rrr}
 0&-1&0\\
 1&0&1\\
 0&-1&0
 \ea
 \right )
 \een
and in  allowing the perceivably less puzzling indexing of the
matrix elements by the parity-symmetric integers
$k=\ldots,-2,-1,0,1,2,\ldots$.

\subsection{$M=1$}

At $M=1$ the set of matching conditions involves just the
following three rows of the central subset of the complete
Schr\"{o}dinger equation $H\psi=E\psi$,
 \ben
\left [\begin {array}{rcccr}
    -1&2\cos \varphi&-1-x&0&0
 \\
 0&-1+x&2\cos \varphi&-1+x&0
  \\{}
 0&0&  -1-x&2\cos \varphi&-1
 \end {array}\right ]\,
  \left [\begin {array}{c} e^{-2{\rm i}\varphi}+
 R\,e^{2{\rm i}\varphi} \\ e^{-{\rm i}\varphi}+
 R\,e^{{\rm i}\varphi}
 \\
 \psi_0
 \\
  T\,e^{{\rm i}\varphi } \\
  T\,e^{2{\rm i}\varphi }
 \end {array}\right ]=0
 \,.
 \een
From the first and third row we get $1+R=(1+x)\psi_0=T$ so that
the remaining row multiplied by $1+x$, viz, equation
 \ben
 (x^2-1)\,( e^{-{\rm i}\varphi}-e^{{\rm i}\varphi}+
 T\,e^{{\rm i}\varphi})+2T\,\cos \varphi+
  (x^2-1)\,T\,e^{{\rm i}\varphi } =0\,.
  \een
leads to the solution in closed form,
 \ben
 T=
 \frac{1}{1+{\rm i}A}\,,
 \ \ \ \ \ \
 R=
 \frac{-{\rm i}A}{1+{\rm i}A}\,,
 \ \ \ \ \ \
 A=\frac{x^2}{1-x^2}\,\cot \varphi\,.
 \een
We may immediately verify that
 \ben
 |R|^2+|T|^2 =1\,.
 \een
This enables us to conclude that in spite of its non-Hermiticity,
our scattering model conserves the probability at $M=1$.

\subsection{$M=2$}

At the next integer index $M=2$ the set of matching conditions
comprises the following five items,
 \ben
\left [\begin {array}{rcccccr}
    -1&2\cos \varphi&-1-x&0&0&0&0
 \\
 0&-1+x&2\cos \varphi&-1&0&0&0
  \\{}
 0&0&  -1&2\cos \varphi&-1&0&0
 \\
 0&0&0&-1&2\cos \varphi&-1+x&0
 \\
 0&0&0&0&-1-x&2\cos \varphi&-1
 \end {array}\right ]\,
  \left [\begin {array}{c}
   e^{-3{\rm i}\varphi}+
 R\,e^{3{\rm i}\varphi} \\ e^{-2{\rm i}\varphi}+
 R\,e^{2{\rm i}\varphi} \\ e^{-{\rm i}\varphi}+
 R\,e^{{\rm i}\varphi}
 +
 \chi_{-1}
 \\
 \psi_0
 \\
  T\,e^{{\rm i}\varphi }+\chi_1 \\
  T\,e^{2{\rm i}\varphi } \\
  T\,e^{3{\rm i}\varphi }
 \end {array}\right ]=0
 \,.
 \een
From the first and last line we get
 \ben
 (1+x)\,\chi_{-1}=-x\,(e^{-{\rm i}\varphi}
 +R\,e^{{\rm i}\varphi})\,,\ \
 \ \ \ \ \
 (1+x)\,\chi_{1}=-x\,T\,e^{{\rm i}\varphi}\,.
 \een
This enables us to consider just the three modified matching
conditions
 \ben
\left [\begin {array}{ccccc}
     -1+x^2&2\cos \varphi&-1&0&0
  \\{}
 0&  -1&2\cos \varphi&-1&0
 \\
 0&0&-1&2\cos \varphi&-1+x^2
  \end {array}\right ]\,
  \left [\begin {array}{c}
   e^{-2{\rm i}\varphi}+
 R\,e^{2{\rm i}\varphi} \\
  e^{-{\rm i}\varphi}+
 R\,e^{{\rm i}\varphi}
 \\
 (1+x)\, \psi_0
 \\
  T\,e^{{\rm i}\varphi } \\
  T\,e^{2{\rm i}\varphi }
 \end {array}\right ]=0
 \,.
 \een
The first row gives
 \ben
 (1+x)\psi_0=1+x^2\, e^{-2{\rm i}\varphi}
 +
 (1+x^2\,e^{2{\rm i}\varphi })\,R\,
 \een
while the third row offers
 \ben
 (1+x)\psi_0=(1+x^2\,e^{2{\rm i}\varphi })\,T\,
 \een
so that we may eliminate $\psi_0$ and obtain the first rule for
$R$ and $T$,
 \ben
 T=R+
 \frac{1+x^2\,e^{-2{\rm i}\varphi }}
 {1+x^2\,e^{2{\rm i}\varphi }}
 =R+\frac{1-{\rm i}\lambda}{1+{\rm i}\lambda}\,,
 \ \ \ \ \lambda=\frac{x^2\sin 2\varphi}{1+x^2\cos 2\varphi}
 \,.
 \een
The remaining middle row leads to the third independent formula
for
 \ben
 (1+x)\psi_0=\frac{1+(R+T)e^{2{\rm i}\varphi }}
 {1+e^{2{\rm i}\varphi }}\,.
 \een
We may combine all three representations of $(1+x)\psi_0$ and
extract the second rule for $R$ and $T$. In the light of the above
representation of the difference $T-R$ we shall complement it by
the second rule which determines the sum $R+T$. Such a recipe
leads to the particularly compact final result,
 \ben
 2\,R=\frac{1-{\rm i}\alpha}{1+{\rm i}\alpha}-
 \frac{1-{\rm i}\beta}{1+{\rm i}\beta}\,,
 \een
 \ben
 2\,T=\frac{1-{\rm i}\alpha}{1+{\rm i}\alpha}+
 \frac{1-{\rm i}\beta}{1+{\rm i}\beta}\,,
 \een
where
 \ben
 \alpha=\frac{x^2\cos 2\varphi \cot \varphi}{1-2x^2\cos^2\varphi}
 \,,
 \ \ \ \ \ \
 \beta=\frac{\sin 2\varphi}{1+x^2\cos 2\varphi}\,.
 \een
Since both $\alpha$ and $\beta$ are real, it is immediate to prove
that
 \ben
 |R|^2+|T|^2 =1\,.
 \een
We see that in the model with $M=2$ the flow of probability is
conserved as well. One feels tempted to expect such a unitary-type
behavior of the amplitudes at all the integer ``interaction
distances" $M$.

Let us test such a conjecture on the next version of our model.

\subsection{$M=3$}

Let us abbreviate $U_{-m}= e^{-m{\rm i}\varphi}+ R\,e^{m{\rm
i}\varphi}$ and $L_n=T\,e^{n{\rm i}\varphi }$ and partition the
seven matching conditions at  $M=3$ as follows,
 \ben
\left [\begin {array}{cc|c|cc} \hline
    2\cos \varphi&-1-x&{}&{}&{}{}
 \\
 {}-1+x&2\cos \varphi&-1&{}&{}{}
  \\
  \hline
 {}{}& \ddots&\ddots&\ddots&{}{}
 \\
 \hline
 {}{}{}&{}&-1&2\cos \varphi&-1+x
 \\
 {}{}{}&{}&{}&-1-x&2\cos \varphi\\
 \hline
 \end {array}\right ]\,
  \left [\begin {array}{c}
  \hline
  U_{-3}\\
  U_{-2}
  +
 \chi_{-2} \\
 \hline
  U_{-1}
 +
 \chi_{-1}
 \\
 \psi_0
 \\
  L_1+\chi_1 \\
  \hline
  L_2+
 \chi_{2} \\
  L_3\\
  \hline
 \end {array}\right ]=
  \left [\begin {array}{c}
  \hline
  U_{-4}\\
  0\\
  \hline 0\\ 0\\ 0\\
  \hline 0\\
  L_4\\
  \hline
 \end {array}\right ]
 \,.
 \een
The first and last lines give
 \ben
 (1+x)\chi_{-2}=-xU_{-2}\,,
 \ \ \ \ \
 (1+x)\chi_{2}=-xL_{2}\,
 \een
and the elimination of the left-hand-side expressions gives the
following reduced set of the five matching conditions,
 \ben
\left [\begin {array}{ccccc} \hline
    2\cos \varphi&-1&{}&{}&{}{}
 \\
 {}-1&2\cos \varphi&-1&{}&{}{}
  \\
 {}{}& -1&2\cos \varphi&-1&{}{}
 \\
 {}{}{}&{}&-1&2\cos \varphi&-1
 \\
 {}{}{}&{}&{}&-1&2\cos \varphi\\
 \hline
 \end {array}\right ]\,
  \left [\begin {array}{c}
  \hline
  U_{-2}
   \\
 (1+x) (U_{-1}
 +
 \chi_{-1})
 \\
 (1+x)\psi_0
 \\
  (1+x)(L_1+\chi_1 )\\
  L_2\\
  \hline
 \end {array}\right ]=
  \left [\begin {array}{c}
  \hline
 (1-x^2) U_{-3}\\
  0\\
   0\\
  0\\
  (1-x^2)L_3\\
  \hline
 \end {array}\right ]
 \,.
 \een
From the first and last equation we eliminate
 \ben
 (1+x)\,\chi_{-1}=-x\,U_{-1}+x^2U_{-3}\,,\ \
 \ \ \ \ \
 (1+x)\,\chi_{1}=-x\,L_1+x^2L_3\,
 \een
and insert these expressions in the remaining three equations,
with the result
 \ben
\left [\begin {array}{ccccc}
     -1&2\cos \varphi&-1&0&0
  \\{}
 0&  -1&2\cos \varphi&-1&0
 \\
 0&0&-1&2\cos \varphi&-1
  \end {array}\right ]\,
  \left [\begin {array}{c}
  U_{-2}\\
 U_{-1}+x^2U_{-3}\\
 (1+x)\psi_0\\
 L_1+x^2L_3\\
 L_2
 \end {array}\right ]=0
 \,.
 \een
Let us rewrite these equations again as the three non-equivalent
definitions of $\psi_0$,
 \ben
 (1+x)\psi_0=U_0+2\,x^2\,\cos \varphi\,U_{-3}\,,
 \een
 \ben
 (1+x)\psi_0=L_0+2\,x^2\,\cos \varphi\,L_{3}\,,
 \een
 \ben
 (1+x)\psi_0=\frac{1}{2\,\cos \varphi}\,
 \left [
 L_1+x^2\,L_{3}+
 U_{-1}+x^2\,U_{-3}
 \right ]
   \een
and eliminate $\psi_0$ in two alternative ways which define the
difference
 \ben
 T-R=\frac{1+2\,x^2\,e^{-3{\rm i}\varphi }\cos \varphi}
 {1+2\,x^2\,e^{3{\rm i}\varphi }\cos \varphi}
 =\frac{1-{\rm i}\gamma}{1+{\rm i}\gamma}\,,
 \een
and the sum
 \ben
 T+R=-e^{-2{\rm i}\varphi }\,
 \frac{1-e^{{\rm i}\varphi }\cos \varphi
 -x^2e^{-2{\rm i}\varphi }\cos 2\varphi}
 {1-e^{-{\rm i}\varphi }\cos \varphi
 -x^2e^{2{\rm i}\varphi }\cos 2\varphi}\,.
 \een
From these formulae it is again easy to derive
 \ben
 |R|^2+|T|^2 =1\,
 \een
i.e., the desirable conservation-of-probability law at $M=3$.

\subsection{$M=4$}

Out of the nine lines of the $M=4$ matching conditions
 \ben
\left [\begin {array}{cc|c|cc} \hline
    2\cos \varphi&-1-x&{}&{}&{}{}
 \\
 {}-1+x&2\cos \varphi&-1&{}&{}{}
  \\
  \hline
 {}{}& \ddots&\ddots&\ddots&{}{}
 \\
 \hline
 {}{}{}&{}&-1&2\cos \varphi&-1+x
 \\
 {}{}{}&{}&{}&-1-x&2\cos \varphi\\
 \hline
 \end {array}\right ]\,
  \left [\begin {array}{c}
  \hline
  U_{-4}\\
  U_{-3}
  +
 \chi_{-3} \\
 \hline
   U_{-2}
 +
 \chi_{-2} \\
  U_{-1}
 +
 \chi_{-1}
 \\
 \psi_0
 \\
  L_1+\chi_1 \\
  L_2+
 \chi_{2} \\
  \hline
  L_3+
 \chi_{3} \\
  L_4\\
  \hline
 \end {array}\right ]=
  \left [\begin {array}{c}
  \hline
  U_{-5}\\
  0\\
  \hline 0\\ 0\\ 0\\ 0\\ 0\\
  \hline 0\\
  L_5\\
  \hline
 \end {array}\right ]
 \,.
 \een
we may eliminate the first and last line using the general formula
 \ben
 (1+x)\chi_{1-M}=-xU_{1-M}\,,
 \ \ \ \ \
 (1+x)\chi_{M-1}=-xL_{M-1}\,.
 \een
Also the rest of the solution can be perceived as a guide to the
construction of the amplitudes $R$ and $T$ at any higher $M$.
Indeed, once we return to the remaining seven matching conditions
at $M=4$,
 \ben
\left [\begin {array}{ccccc} \hline
    2\cos \varphi&-1&{}&{}&{}{}
 \\
 {}-1&2\cos \varphi&-1&{}&{}{}
  \\
 {}{}& \ddots&\ddots&\ddots&{}{}
 \\
 {}{}{}&{}&-1&2\cos \varphi&-1
 \\
 {}{}{}&{}&{}&-1&2\cos \varphi\\
 \hline
 \end {array}\right ]\,
  \left [\begin {array}{c}
  \hline
  U_{-3}
   \\
 (1+x) (U_{-2}
 +
 \chi_{-2})\\
 (1+x) (U_{-1}
 +
 \chi_{-1})
 \\
 (1+x)\psi_0
 \\
  (1+x)(L_1+\chi_1 )\\
  (1+x)(L_2+\chi_2 )\\
  L_3\\
  \hline
 \end {array}\right ]=
  \left [\begin {array}{c}
  \hline
 (1-x^2) U_{-4}\\
  0\\
   0\\
  0\\
   0\\
  0\\
  (1-x^2)L_4\\
  \hline
 \end {array}\right ]
 \,
 \een
we may repeat the algorithm and eliminate its first and last line.
Another general pair of formulae serves the purpose,
 \ben
 (1+x)\,\chi_{2-M}=-x\,U_{2-M}+x^2U_{-M}\,,\ \
 \ \ \ \ \
 (1+x)\,\chi_{M-2}=-x\,L_{M-2}+x^2L_M\,
 \een
after one inserts $M=4$. In the subsequent step of the reduction
procedure we arrive at the quintuplet of equations
 \ben
\left [\begin {array}{ccccc} \hline
    2\cos \varphi&-1&{}&{}&{}{}
 \\
 {}-1&2\cos \varphi&-1&{}&{}{}
  \\
 {}{}& -1&2\cos \varphi&-1&{}{}
 \\
 {}{}{}&{}&-1&2\cos \varphi&-1
 \\
 {}{}{}&{}&{}&-1&2\cos \varphi\\
 \hline
 \end {array}\right ]\,
  \left [\begin {array}{c}
  \hline
  U_{-2}+\chi_{-2}
   \\
 U_{-1}
 +
 \chi_{-1}
 \\
 \psi_0
 \\
  L_1+\chi_1 \\
  L_2+\chi_2\\
  \hline
 \end {array}\right ]=
  \left [\begin {array}{c}
  \hline
 U_{-3}/(1+x)\\
  0\\
   0\\
  0\\
  L_3/(1+x)\\
  \hline
 \end {array}\right ]
 \,.
 \een
Using the first and fifth equation again, we specify the last
auxiliary quantities.
 \ben
 (1+x)\,\chi_{-1}=-x\,U_{-1}+2x^2\cos \varphi\,U_{-4}\,,\ \
 \ \ \ \ \
 (1+x)\,\chi_{1}=-x\,L_1+2x^2\cos \varphi\,L_4\,.
 \een
This exemplifies the last step of the generic recurrent recipe
because the next step will already involve the exceptional central
element $\psi_0$. Thus, our knowledge of the expressions for
$\chi_{\pm 1}$ leads to the final triplet of conditions
 \ben
 (1+x)\psi_0=U_0+x^2(1+2\,\cos 2\varphi)U_{-4}\,,
 \een
 \ben
 (1+x)\psi_0=L_0+x^2(1+2\,\cos 2\varphi)L_{4}\,,
 \een
 \ben
 (1+x)\psi_0=
 \frac{L_1+U_{-1}}{2\,\cos \varphi}+x^2(L_4+U_{-4})\,.
  \een
After the two alternative eliminations of $\psi_0$ we routinely
arrive at our last two linear equations for the two unknown
quantities $R+T$ and $T-R$. Their elementary though a bit clumsy
solution will not be displayed here anymore. Whenever asked for,
the proof of the conservation law at $M=4$ as well as the further,
more or less routine though increasingly tedious continuation of
our construction to the higher ``distances $M$ between
interactions" are left to the readers.

\section{Summary}

One of the most pleasant and encouraging observations made during
many practical applications of quantum theory is that our basic
understanding of experimental data can often be provided by fairly
elementary mathematical models. Among them, a prominent role is
played by the one-dimensional Schr\"{o}dinger equation. Of course,
the detailed physical interpretation of such a class of models can
vary with the experimental setup and may range from the naive
fitting scenario up to a schematic reduction of field theory to
zero dimensions.

In the latter, highly speculative context Bender and Milton
\cite{BM} and Bender and Boettcher \cite{BB} revealed that
phenomenological as well as theoretical purposes could be served
very well by complex potentials exemplified by $V(x) = {\rm i}x^3$
and supporting real spectra of bound states \cite{DDT}. Later on,
it has been clarified that the transition to the complex $V(x)$
does not in fact violate any rules of Quantum Mechanics because
even for complex potentials the Hamiltonian can be reinterpreted
as self-adjoint after a suitable adaptation of the Hilbert space
of states \cite{Ali}.

Jones \cite{Jones} was probably the first author who analyzed the
possibilities of the same adaptation of the Hilbert space in the
scattering scenario. Although he choose one of the simplest and
best understood potentials, viz., the delta function with a
complex coupling, his conclusions concerning both the mathematical
feasibility and the physical clarity of the complexified
scattering problem were rather discouraging. His construction
revealed that in spite of the ultralocal form of his toy model the
scattered waves proved perceivably and counterintuitively
distorted.

In our present note we reanalyzed the situation by incorporating,
in explicit manner, the postulate of the so called ${\cal
PT}-$symmetry of the Hamiltonian which is often being implemented
in the constructive description of bound states in unusual Hilbert
spaces. For this purpose we introduced and solved and entirely new
class of discrete models of scattering. We were really surprised
when we revealed that these models behaved {\em differently} in
comparison with their similar ${\cal PT}-$asymmetric predecessors
of refs.~\cite{Jones,prd}.

The key merit of our present family of models should be seen in
the fact that not quite expectedly, they fully conserve the
probability and do not seem to exhibit any signs of an asymptotic
non-locality. Moreover, since they are simple and exactly
solvable, the emerging possibilities of their entirely standard
practical applications and/or theoretical probabilistic
interpretation do not seem to be an artifact of their present
discretized mathematical form.

We believe that on the background of certain pessimistic
physics-related perspectives as formulated in
refs.~\cite{Jones,prd}, our present results could serve as a
source of new optimism, needed for the continuation of the search
for some new manifestly non-Hermitian models of scattering. One
can hope that the  user-friendly features of our models will
survive their extensions, both in the sense of returning to the
continuous limit $h \to 0$ {\em and} in the sense of finding their
more-parametric descendants of a greater descriptive flexibility.

 \vspace{5mm}

\section*{Acknowledgement}

Work supported by the M\v{S}MT ``Doppler Institute" project Nr.
LC06002,  by the Institutional Research Plan AV0Z10480505 and by
the GA\v{C}R grant Nr. 202/07/1307.



\end{document}